\pdfoutput=1
\documentclass[a4paper]{jpconf}
\newcommand{\MJD}{\textsc{Majorana Demonstrator}}
\usepackage{graphicx}

\begin{document}
\title{Data quality assurance for the \sc{Majorana Demonstrator}}
\author{
J Myslik$^1$,
N Abgrall$^1$,
S I Alvis$^2$,
I J Arnquist$^3$,
F T Avignone~III$^{4,5}$,
A S Barabash$^6$,
C J Barton$^7$,
F E Bertrand$^5$,
T Bode$^8$,
A W Bradley$^1$,
V Brudanin$^9$,
M Busch$^{10,11}$,
M Buuck$^2$,
T S Caldwell$^{12,11}$,
Y-D Chan$^1$,
C D Christofferson$^{13}$,
P-H Chu$^{14}$,
C Cuesta$^{2,15}$
J~A~Detwiler$^2$,
C~Dunagan$^{13}$,
Yu Efremenko$^{16,5}$,
H Ejiri$^{17}$,
S~R~Elliott$^{14}$,
T Gilliss$^{12,11}$,
G~K~Giovanetti$^{18}$,
M P Green$^{19,11,5}$,
J~Gruszko$^2$,
I S Guinn$^2$,
V~E~Guiseppe$^4$,
C R Haufe$^{12,11}$,
L Hehn$^{1}$,
R~Henning$^{12,11}$,
E~W~Hoppe$^3$,
M A Howe$^{12,11}$,
K J Keeter$^{20}$,
M~F~Kidd$^{21}$,
S~I~Konovalov$^6$,
R T Kouzes$^3$,
A M Lopez$^{16}$,
R D Martin$^{22}$,
R~Massarczyk$^{14}$,
S J Meijer$^{12,11}$,
S Mertens$^{8,23}$,
C O'Shaughnessy$^{12,11}$,
G Othman$^{12,11}$,
W Pettus$^{2}$,
A W P Poon$^1$,
D C Radford$^5$,
J~Rager$^{12,11}$,
A L Reine$^{12,11}$,
K Rielage$^{14}$,
R G H Robertson$^2$,
N~W~Ruof$^2$,
B~Shanks$^{5}$,
M Shirchenko$^9$,
A M Suriano$^{13}$,
D~Tedeschi$^4$,
J~E~Trimble$^{12,11}$,
R L Varner$^5$,
S Vasilyev$^9$,
K Vetter$^{1,24}$,
K~Vorren$^{12,11}$,
B R White$^{14}$,
J F Wilkerson$^{12,11,5}$,
C Wiseman$^4$,
W Xu$^7$,
E~Yakushev$^9$,
C-H Yu$^5$,
V Yumatov$^6$,
I Zhitnikov$^9$
and
B X Zhu$^{14}$
}

\address{$^1$ Nuclear Science Division, Lawrence Berkeley National Laboratory, Berkeley, CA, USA}
\address{$^2$ Center for Experimental Nuclear Physics and Astrophysics, and Department of Physics, University of Washington, Seattle, WA, USA}
\address{$^3$ Pacific Northwest National Laboratory, Richland, WA, USA}
\address{$^4$ Department of Physics and Astronomy, University of South Carolina, Columbia, SC, USA}
\address{$^5$ Oak Ridge National Laboratory, Oak Ridge, TN, USA}
\address{$^6$ National Research Center ``Kurchatov Institute'' Institute for Theoretical and Experimental Physics, Moscow, Russia}
\address{$^7$ Department of Physics, University of South Dakota, Vermillion, SD, USA} 
\address{$^8$ Max-Planck-Institut f\"{u}r Physik, M\"{u}nchen, Germany}
\address{$^9$ Joint Institute for Nuclear Research, Dubna, Russia}
\address{$^{10}$ Department of Physics, Duke University, Durham, NC, USA}
\address{$^{11}$ Triangle Universities Nuclear Laboratory, Durham, NC, USA}
\address{$^{12}$ Department of Physics and Astronomy, University of North Carolina, Chapel Hill, NC, USA}
\address{$^{13}$ South Dakota School of Mines and Technology, Rapid City, SD, USA}
\address{$^{14}$ Los Alamos National Laboratory, Los Alamos, NM, USA}
\address{$^{15}$ Present address:
Centro de Investigaciones Energ\'{e}ticas, Medioambientales y Tecnol\'{o}gicas, CIEMAT, 28040, Madrid, Spain}
\address{$^{16}$ Department of Physics and Astronomy, University of Tennessee, Knoxville, TN, USA}
\address{$^{17}$ Research Center for Nuclear Physics, Osaka University, Ibaraki, Osaka, Japan}
\address{$^{18}$ Department of Physics, Princeton University, Princeton, NJ, USA}
\address{$^{19}$ Department of Physics, North Carolina State University, Raleigh, NC, USA}	
\address{$^{20}$ Department of Physics, Black Hills State University, Spearfish, SD, USA}
\address{$^{21}$ Tennessee Tech University, Cookeville, TN, USA}
\address{$^{22}$ Department of Physics, Engineering Physics and Astronomy, Queen's University, Kingston, ON, Canada} 
\address{$^{23}$ Physik Department, Technische Universit\"{a}t, M\"{u}nchen, Germany}
\address{$^{24}$ Alternate address: Department of Nuclear Engineering,
University of California, Berkeley, CA, USA}

\ead{jwmyslik@lbl.gov}

\begin{abstract}
The \textsc{Majorana Demonstrator} is an experiment constructed to search for
    neutrinoless double-beta decays in germanium-76 and to demonstrate the
    feasibility to deploy a large-scale experiment in a phased and modular
    fashion. It consists of two modular arrays of natural and
    $^{76}$Ge-enriched germanium detectors totalling 44.1 kg, located at the
    4850' level of the Sanford Underground Research Facility in Lead, South
    Dakota, USA. Any neutrinoless double-beta decay search requires a thorough
    understanding of the background and the signal energy spectra. The various 
    techniques employed to ensure the integrity of the
    measured spectra are discussed. Data collection is monitored with a thorough set of
    checks, and subsequent careful analysis is performed to qualify the
    data for higher level physics analysis. Instrumental background events
    are tagged for removal, and problematic channels are removed from consideration as necessary.
\end{abstract}

\section{Introduction}
The process of two neutrino 
double-beta decay ($(A,Z)\rightarrow(A,Z+2) + 2e^{-} + 2\bar{\nu}_{e}$) is
lepton number conserving, and has been observed in multiple isotopes
for which a beta decay is energetically forbidden (e.g. $^{76}$Ge, $^{130}$Te,
and $^{136}$Xe).  However, the process of neutrinoless double-beta decay,
($(A,Z)\rightarrow(A,Z+2) + 2e^{-}$) violates lepton number conservation, and
has not yet been observed.  As some theoretical models that give neutrinos mass
require lepton number violation, searching for it is therefore of experimental
interest. 

The main physics purpose of the \MJD{} \cite{Abgrall:2013rze} is the search for lepton number violation
 through the process of neutrinoless double-beta decay.  While the sum of the
 energies of the electrons produced in a two neutrino double-beta decay is a
 continuum, neutrinoless double-beta decay produces electrons with energies summing to the
 Q-value of the decay (2039~keV for $^{76}$Ge).  This results in a signature that
 can be searched for experimentally, though good energy resolution is
 important for separating the neutrinoless double-beta decay signal from the
 irreducible background of the two neutrino double-beta decay spectrum.
 Ultra-low backgrounds are also important to such a rare event search, as is
 the ability to scale up to a larger mass detector, to improve the probability
 of observing a decay in a shorter period of time.

 High-purity germanium detectors have a number of advantages for a
 neutrinoless double-beta decay search.  Their excellent energy resolution
 provides excellent separation between the zero neutrino signal and the
 two neutrino background.  Also, the germanium crystals can be made from
 material enriched in $^{76}$Ge.  This integrates the double-beta decay isotope
 into the detector active volume without producing any challenges for detector
 performance.

Therefore, the \MJD{} has 3 design goals:
 \begin{itemize}
    \item Demonstrating backgrounds low enough to justify building a tonne
        scale experiment.
     \item Establishing feasibility to construct and field modular arrays of Ge
         detectors.
     \item Searching for additional physics beyond the Standard Model. 

 \end{itemize}

The \MJD{} is currently operating on the 4850' level of the Sanford Underground
Research Facility (SURF) in Lead, South Dakota, USA.  It consists of 2
cryostat modules, each containing 7 strings of 3 to 5 p-type Point-Contact (PPC)
germanium crystal detectors.  The longer drift time of PPC detectors helps
distinguish signal-like single-site interactions from multi-site background
interactions.  The total mass of the detectors is 44.1~kg, of which 14.4~kg are
natural germanium crystals, and 29.7~kg are enriched to 88\% $^{76}$Ge.
These cryostat modules are surrounded by copper then
lead to shield against gamma rays, active muon veto panels, then polyethyelene to
shield against neutrons. The boundary between the lead and the muon veto panels 
is an aluminum enclosure that is sealed and purged with liquid nitrogen
boil-off gas to keep radon away from the cryostats. 
In addition, using radiopure materials in the construction of the
experiment was a high priority, in order to achieve the ultra-low backgrounds
necessary for such a rare event search \cite{Abgrall:2016cct}. The ultra-low
backgrounds of the \MJD{},
combined with its excellent energy resolution, also make it a multipurpose
detector, capable of searching for additional physics beyond the Standard Model
\cite{Abgrall:2016tnn}.

In a rare event search experiment with ultra-low backgrounds, events
originating from hardware abnormalities in the detector instrumentation may threaten to
dominate the backgrounds for the experiment.  Therefore, data quality is
monitored during data collection, specific instrumental background event
pathologies are identified and either corrected, tagged for removal, or handled
in later analysis, and further in-depth analyses
performed in order to ensure clean data for physics analysis.  The following
sections describe these data quality assurance efforts in more detail.

\section{Monitoring of data collection}
Ensuring good data quality requires continuously being vigilant for problems
with the data taking.  With this in mind, the interface to the slow controls database
(a CouchDB~\cite{anderson2010couchdb} database) contains information on
detector conditions in real time.  This includes event rates and detector baselines,
along with hardware status information (e.g. liquid nitrogen fill levels), and
lab environmental conditions (e.g. temperature, humidity, and particulate count).  Further processing is also conducted
onsite, to allow for near-term monitoring of the data collected.

Once a run is complete ($\sim 1$ hour during normal data taking) and on the
local RAID array, the data is transferred to the Parallel Distributed Systems Facility (PDSF) cluster at the National Energy Research Scientific
Computing Center (NERSC), where it is processed within approximately 30 minutes.  Metadata for these runs is then
uploaded to the run database (also CouchDB).  The metadata includes checks of the DAQ
configuration and the run conditions as they relate to data quality, such as
whether the DAQ system is properly configured for physics running, as well as
that there are no indications of other issues.  For example, this includes
ensuring that the run is long enough, the radon purge rate is sufficient, and
the event rate is not too high.  This information is combined to assign a rank
to the run: gold, silver, bronze, or bad.  The run metadata also includes  plots that are made of key indicators of
detector performance, such as the event rates in each channel, and a coarsely
binned energy spectrum as a function of time.  The DAQ shifter examines the slow controls and run
databases four times per day during normal operations, keeping watch for signs
of data quality irregularities in both the live information and the run metadata.

\begin{figure}[!htb]
    \includegraphics[trim=5.2cm 5cm 3.5cm 3cm,clip,width=0.49\textwidth]{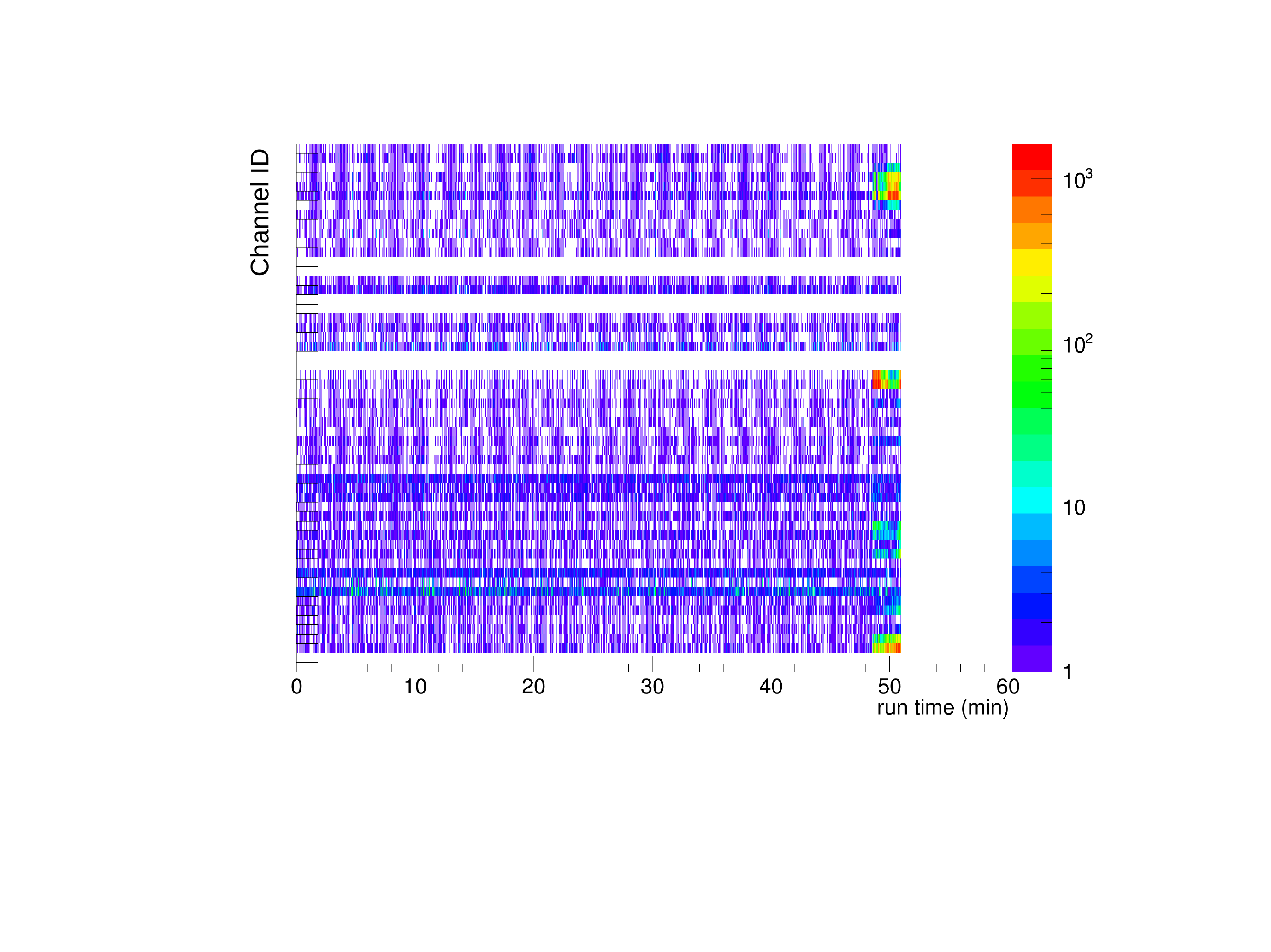}
    \includegraphics[trim=0 0.3cm 0 1cm,clip,width=0.49\textwidth]{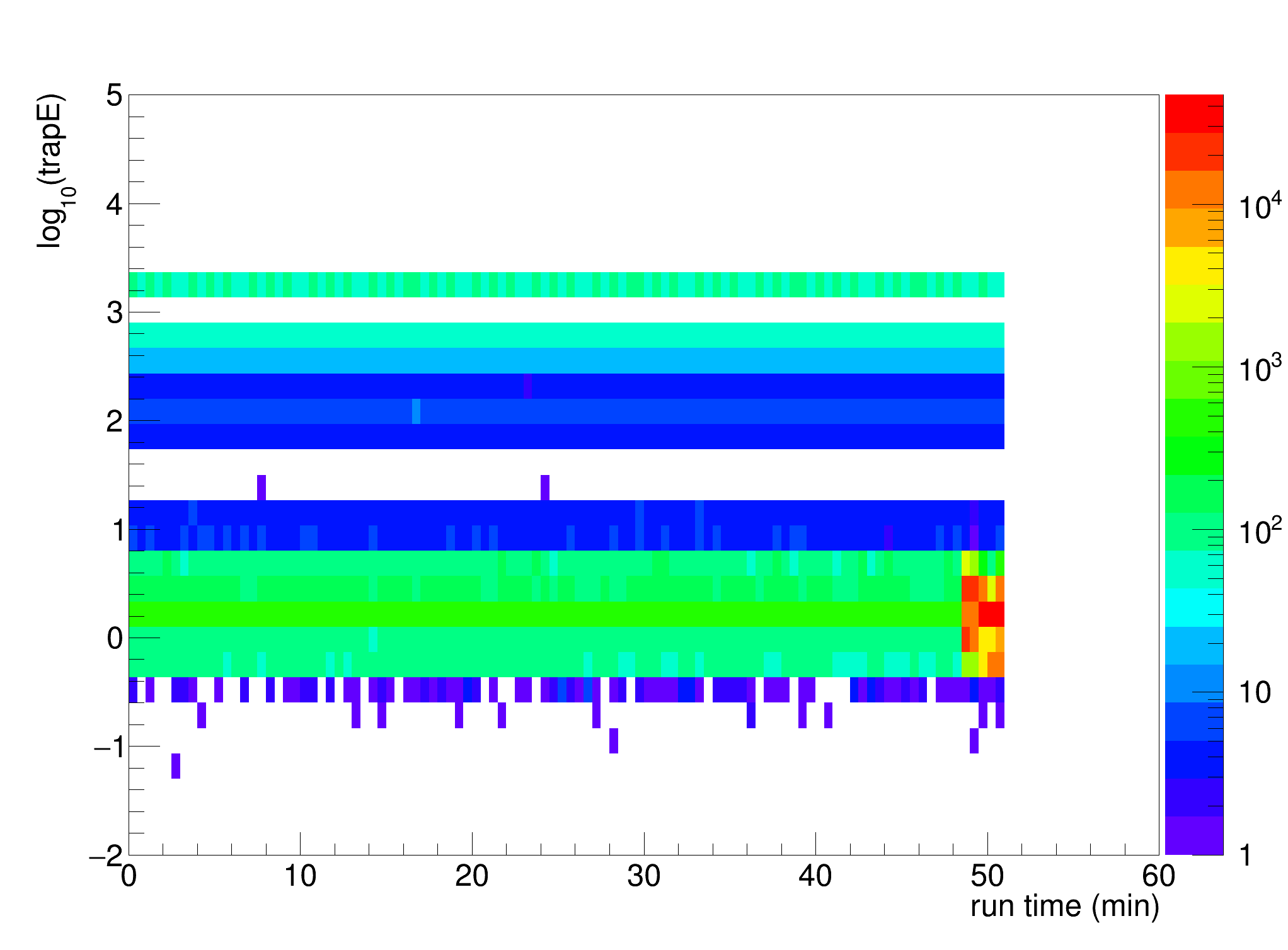}
    \caption{Channel-by-channel rates (left) and the summed energy (right) as a
    function of time in a run, from the run database.  In the
    channel-by-channel rates plot, channels are sorted by their position in the
    module and whether they are the high or low gain channel. A liquid nitrogen fill
    produced low energy noise at the end of this run.\label{fig:RunDBPlots}}
\end{figure}

A liquid nitrogen fill is a regular occurrence, and the resulting microphonics
are apparent in the run metadata plots. An example is shown in Figure~\ref{fig:RunDBPlots}.

\section{Removal of instrumental backgrounds}
Instrumental backgrounds originate in the electronics used to read out the
detectors.  This may be due to a hardware abnormality (e.g. a random bit flip in the
ADC), but can also be a simple matter of the suitability of the hardware
configuration for a given event (e.g. the gain for a channel resulting
in a saturated ADC for a high energy event). In some cases, instrumental backgrounds are sufficient to
trigger data taking, and result in pure instrumental background events (e.g. triggering on a positive spike on baseline caused by
a random bit flip in the ADC).  In
other cases, an otherwise  good physics event may be marred by the presence of
an instrumental abnormality (e.g. the positive spike from a random bit flip in
the ADC coincides with a physics waveform).  This could lead to problems or inaccuracies in the event
reconstruction.

\begin{figure}[!htb]
    \includegraphics[trim=0 0 0 1cm, clip,width=0.49\textwidth]{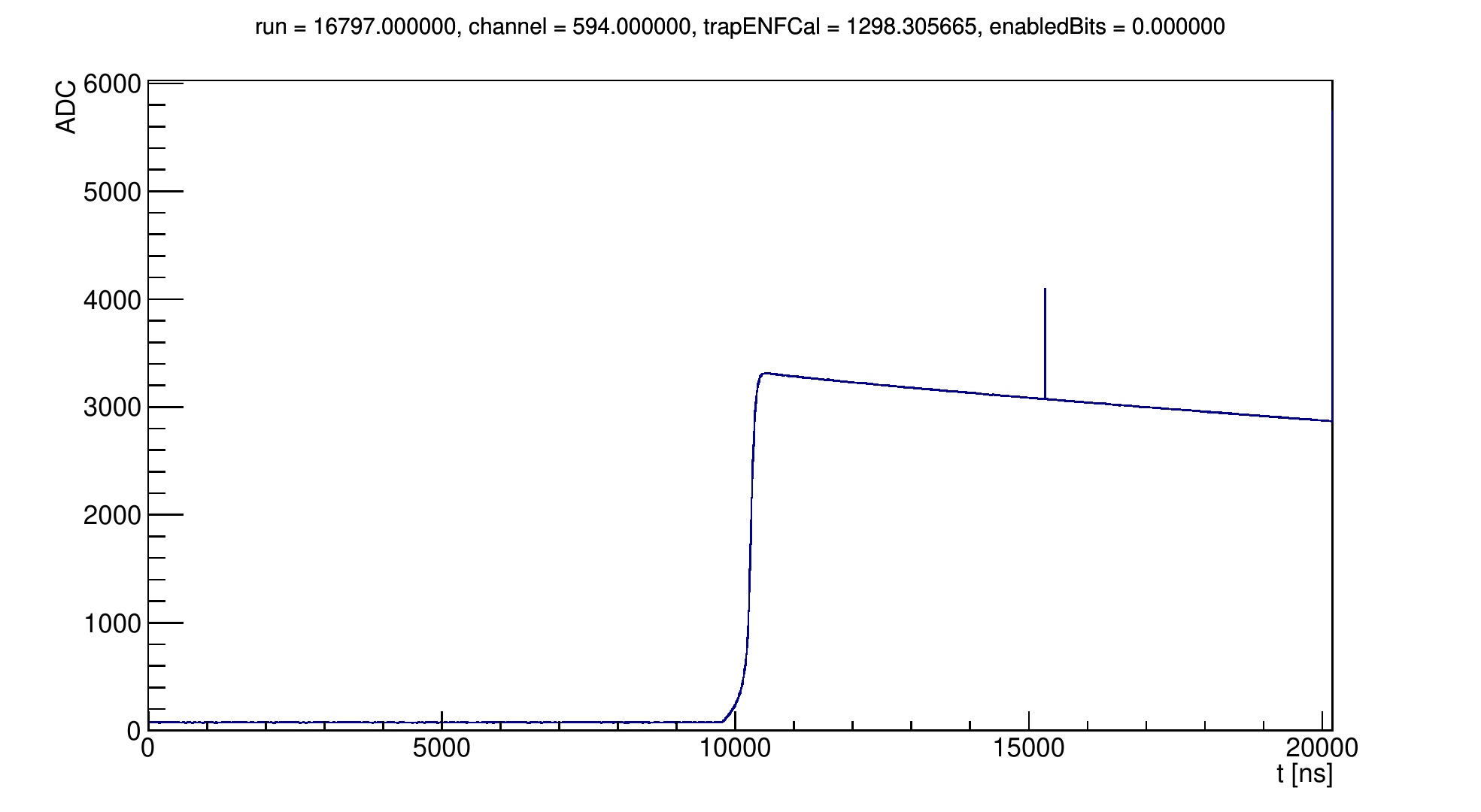}
    \includegraphics[trim=0 0 0 1cm, clip,width=0.49\textwidth]{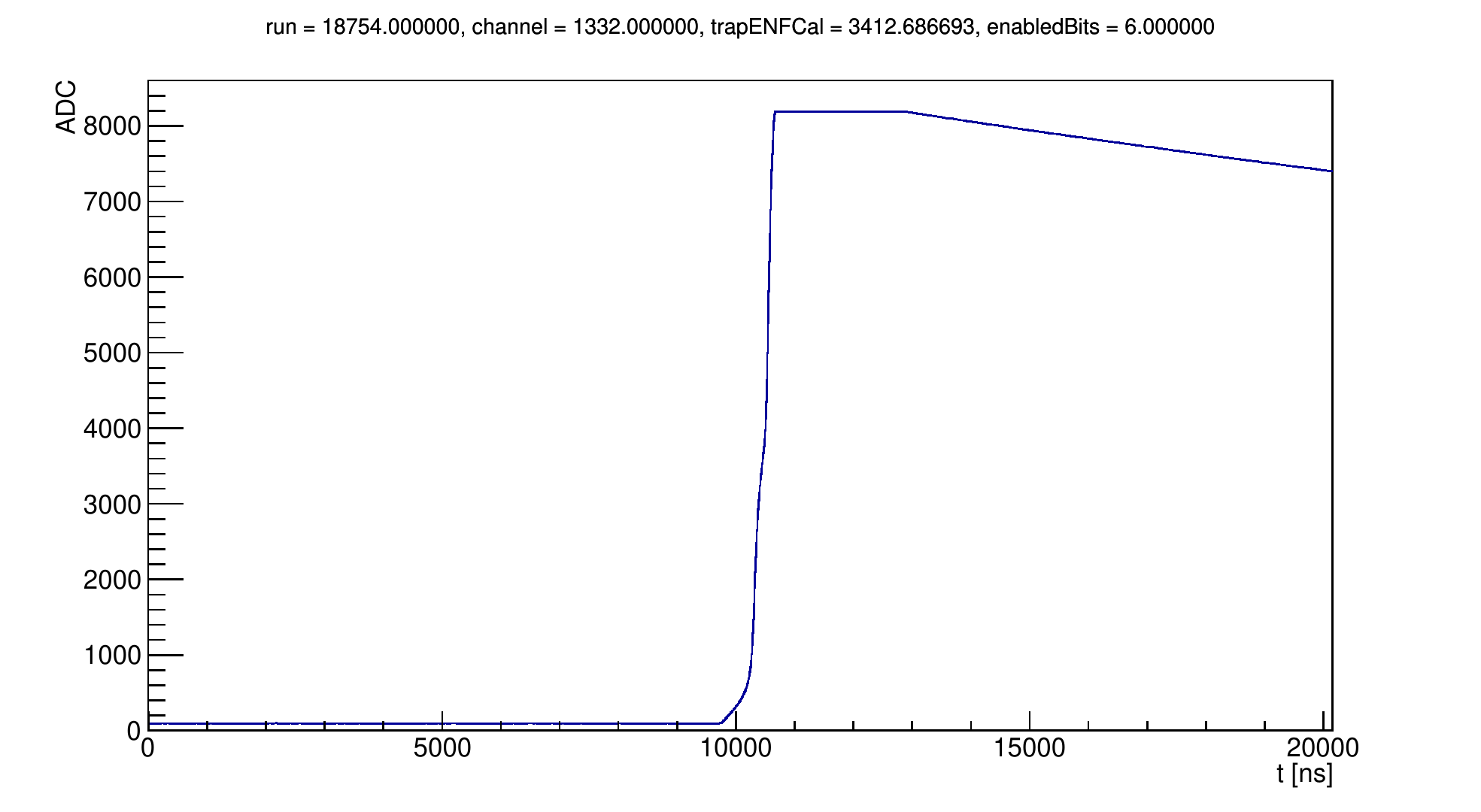}
    \caption{Instrumental background waveforms.  On the left, a random ADC bit
    flip caused a spike on the waveform.  This is identified and corrected
    during data processing.  On the right, the analog signal exceeded the
    maximum ADC value.  These are tagged for replacement with the detector's
    low gain channel waveform. \label{fig:IBWFs}}
\end{figure}

Dedicated analyses are performed during waveform reconstruction in
order to identify instrumental background waveforms.  Once identified, the
instrumental background waveforms can either
be corrected or tagged so that they can be dealt with properly during the
physics analysis.  Correcting an instrumental background waveform can involve
either removing the pathology added to it by the instrumental background, or
replacing the marred waveform with the corresponding waveform in the detector's low gain
channel (which may not have the same problem).  Two examples of instrumental background waveforms, and how
they are dealt with, are shown in Figure~\ref{fig:IBWFs}.  To ensure that no
unexpected instrumental backgrounds affect the final event sample, a visual
scan of waveforms is performed. Ultimately, studies of calibration data and of
the measured two neutrino double-beta decay spectrum indicate that instrumental
backgrounds affect a very small fraction of physical events (less than 0.1\%). 

\section{Selection of data for physics analysis}
The first step in selection of data for physics analysis makes use of the run
metadata.  The rank in the run database provides a quick way to order runs
by their expected data quality, but a final determination on acceptable runs involves
considering the run metadata relative to the detector operating conditions, and checking
the run metadata plots for any anomalies potentially not captured in the run
rank.

Beyond this initial selection of runs, more detailed studies are performed to probe data
quality on a run-by-run and channel-by-channel basis.  The timing of injected
pulses is examined in order to ensure that the channels are synchronized
properly.  Channels that have insufficient good calibration data to be properly calibrated
are flagged for removal.  The stability of computed parameters used in the
physics analysis (e.g. for removal of multi-site background events) is checked
across the dataset, and channels are excluded for runs where detector
performance has caused these runs to deviate too far from their nominal values.
Finally, periods of anomalously high instrumental background rates are
examined, and sufficiently problematic channels are excluded.

Once all of these checks have been performed, the runs and channels that remain are considered
to provide data of sufficient quality for physics analysis. During the analysis
the tagged instrumental background waveforms are removed, as well as events in periods of
time corresponding to a module's liquid nitrogen fills, and any events flagged
by the active muon veto.  These analysis cuts on the good runs and channels
ensure that only high quality data is provided for further physics analysis.

\section{Conclusions}
The \textsc{Majorana Demonstrator} is a neutrinoless double-beta decay
experiment, currently operating on the 4850' level of the Sanford Underground
Research Facility.  Understanding the signal and background energy spectra is
of paramount importance to the experiment, and in this ultra-low background
environment, data quality issues could contribute significantly to these
spectra.  Therefore, between collection of the data and its use for physics analysis
are multiple checks of data quality.  The outcome of these checks is used to determine
which data is of sufficient quality to be analyzed for the neutrinoless double-beta decay search.

\ack
This material is based upon work supported by the U.S. Department of Energy, Office of Science, Office of Nuclear Physics, the Particle Astrophysics and Nuclear Physics Programs of the National Science Foundation, and the Sanford Underground Research Facility.
\section*{References}
\bibliographystyle{iopart-num}
\bibliography{proceedings}

\end{document}